Comment on *Infrared dielectric function of GaAs$_{1-x}$P$_x$ semiconductor alloys near the reststrahlen bands* [Appl. Phys. Lett. **123**, 172102 (2023)]

A. Elmahjoubi, T. Alhaddad, A.V. Postnikov and O. Pagès[a)]
LCP-A2MC, UR 4632, Université de Lorraine, 57000 Metz, France

[a)] Author to whom correspondence should be addressed: olivier.pages@univ-lorraine.fr

Recently, Zollner *et al.*[1] reported on a novel far-infrared (IR) ellipsometry study of the lattice vibrations (phonons) of GaAs$_{1-x}$P$_x$ single crystals deposited on GaAs spanning the composition domain. Their abundant data reveal separate (transverse optical – TO, longitudinal optical – LO) Ga-As (~270 cm$^{-1}$) and Ga-P (~360 cm$^{-1}$) phonon bands, with a fine structure in each, especially clear for the Ga-P band at minor P content. The fine structure has earlier been observed by IR-reflectivity.[2,3] Hence, it is intrinsic to GaAs$_{1-x}$P$_x$ and worth to discuss. So far, two models were proposed to explain this fine structure, *i.e.*, the cluster model[3] (CM) and our percolation model[4] (PM).[1] Zollner *et al.* suggest that their IR-ellipsometry data are qualitatively better explained by the CM than by the PM.[1] Yet, they admit being unable to set a robust CM framework for their data, and question the physical significance of using the CM in this case.[1] On our side, we must admit that our original version of the PM for GaAs$_{1-x}$P$_x$[4] does not accurately reproduce the intensity interplay within the Ga-P fine structure when $x$ varies, as apparent in Zollner's data (detail is given below). Our aim in this comment is to design an improved version of the PM for GaAs$_{1-x}$P$_x$ that remedies such drawback and explains the phonon behavior of GaAs$_{1-x}$P$_x$.

First, we briefly put the subject into perspective, and then illustrate our point with a short discussion, deploying details as supplementary material (prefixed S below).

The CM[3] and the PM (in its original version)[4] assume that the like bonds of a given species vibrate at different frequencies in an alloy depending on their first-neighbor environment (Sec. SI). The latter is described in the real three-dimensional crystal in the CM (1-bond→4-phonon scheme) and at one-dimension (along the linear chain approximation) in the PM (1-bond→2-phonon scheme). Verleur and Barker (VB) achieved a fair modeling of their GaAs$_{1-x}$P$_x$ IR-reflectivity spectra within the CM, albeit only by assuming a far-from-random As↔P substitution.[3] Yet, GaAs$_{1-x}$P$_x$ later proved to be random.[5] Two decades ago some of us simulated the same IR-reflectivity spectra by the PM based on a random substitution.[4] Cebulski *et al.* later operated the CM on the same basis to fit the imaginary part of the relative dielectric function $\varepsilon_r(\omega, x)$ of GaAs$_{1-x}$P$_x$, reconstructed from a Kramers-Kronig transformation of the raw/digitized IR-reflectivity spectra taken by VB.[6] However, a fair fitting by Cebulski *et al.* required an additional (irregular/unassigned) GaP-like phonon mode (APM) besides the four "canonical" (regular) phonon modes (CPM) provided by the CM.[6]

From this brief overview, it emerges that the CM on its own fails to explain *quantitatively* the phonon spectra of random GaAs$_{1-x}$P$_x$. Even *qualitatively*, its use is problematic, as pointed out by Zollner *et al.*[1] (see above). The PM offers an alternative option. Yet, the existing PM for random GaAs$_{1-x}$P$_x$[4] is not exact when confronted to Zollner's data,[1] as already mentioned. Hence, to date, the GaAs$_{1-x}$P$_x$ phonon behavior remains unexplained.

The novel and abundant IR-Ellipsometry data – kindly made available to us by Zollner *et al.*[1] (Fig. S1, symbols), bring motivation to elaborate an improved version of the PM for random GaAs$_{1-x}$P$_x$, our aim in this comment. For doing so, we focus on Zollner's $Im\{\varepsilon_r(\omega, x)\}$ data – in fact remarkably consistent with those of Cebulski *et al.*[6] – obtained by inversion of the measured raw ellipsometry angles through a fitting process that solves the Fresnel equations of the GaAs$_{1-x}$P$_x$/GaAs heterostructure.[1] Such focus is justified since $Im\{\varepsilon_r(\omega, x)\}$ relates to the TO modes.[7] The latter are well suited to address the nature of the phonon behavior of an alloy because they assimilate with purely-mechanical (mass + spring) harmonic oscillators.[2] As such, they hardly couple, and hence preserve the natural diversity of the phonon pattern of a complex system such as an alloy (Sec. SI).

Omitting minor features (marked by asterisks in Fig. S1; eliminated by slightly changing the alloy composition, hence likely due to slight composition inhomogeneities[1,8]), the $Im\{\varepsilon_r(\omega, x)\}$ data of Zollner et al.[1] (Fig. S1, symbols) can be grasped in three {1×(Ga-As),2×(Ga-P)} main lines. A dramatic intensity interplay occurs within the Ga-P doublet. The lower line dominates over the upper one at large $x$ values, yet the relation gets inverted towards small $x$. Intensity matching between the two Ga-P lines is achieved at $x$~0.275 in the experimental $Im\{\varepsilon_r(\omega,x)\}$ data,[1,6] and not at $x$~0.5 as predicted by our original first-neighbor version of the PM.[4] The shift in the above critical $x$ value naturally comes out of the PM, still assuming a random As↔P substitution, provided the sensitivity of Ga-P vibrations to the local environment is extended from first over second neighbors.

An updated second-neighbor version of the PM for random GaAs$_{1-x}$P$_x$ pictures an overall four-mode $\{TO_{Ga-As}^{P,(2)}, TO_{Ga-As}^{As,(2)}, TO_{Ga-P}^{P,(2)}, TO_{Ga-P}^{As,(2)}\}$ pattern – in this order of frequency (Fig. S2). The subscript and double superscript mark the vibrating unit and its local environment (nature, length scale), respectively. The four-TO pattern is supported by *ab initio* (SIESTA) calculations of the TO frequencies at minimum alloy disorder ($x$~0,1 – Fig. S3). The extension from first- over second-neighbors, impacting the TO intensities, is supported by *ab initio* phonon calculations at maximum alloy disorder ($x$=0.5) using large (128-atom and 216-atom) GaAs$_{1-x}$P$_x$ supercells with special quasirandom structures[9] (Fig. S4).

Fair contour modeling (Fig. S1, curves) of the GaAs$_{1-x}$P$_x$ $Im\{\varepsilon_r(\omega,x)\}$ spectra of Zollner et al.[1] (symbols) is achieved by using a compact version of the above generic four-TO PM scheme in which the two Ga-As lines are merged into one (Sec. SII), without free parameter basically.

Hence, the second-neighbor PM scheme proposed in this comment takes a major step towards understanding the phonon behavior of random GaAs$_{1-x}$P$_x$, beyond existing approaches. The latter statement is supported by an overview of the CM and PM schemes so far tested on GaAs$_{1-x}$P$_x$ – focused on the sensitive Ga-P signal at $x$=0.275, that helps to evaluate their relative bias and merits (Fig. S5). Last, we check for completeness how the latter signal is impacted by deviating from the ideally random substitution (Fig. S6). Generally, second-neighbor PM schemes were proposed for Zn-based II-VI alloys.[10-12] GaAs$_{1-x}$P$_x$ is the first III-V case.

**Supplementary material**
Supplementary material covers various issues. The PM is directly tested on the raw GaAs$_{1-x}$P$_x$ $Im\{\varepsilon_r(\omega,x)\}$ data of Zollner et al.[1] (Fig. S1). The used PM version is derived from a generic second-neighbor four-mode {2×(Ga-As),2×(Ga-P)} PM scheme for disordered GaAs$_{1-x}$P$_x$ (Fig. S2) based on *ab initio* phonon calculations at minimum ($x$~0,1 – frequency aspect, Fig. S3) and maximum ($x$=0.5 – intensity aspect, Fig. S4) disorder. The comparative merits of the CM and of the PM to describe the sensitive Ga-P phonon pattern of GaAs$_{1-x}$P$_x$ are discussed by focusing on $x$=0.275 (Fig. S5). Last, we test how of the $Im\{\varepsilon_r(\omega, x = 0.275)\}$ GaP-like signal is impacted within the PM by deviating from the ideally random As↔P substitution (Fig. S6).

**Acknowledgements**
We thank S. Zollner et al.[1] for sharing with us their abundant raw $Im\{\varepsilon_r(\omega,x)\}$ data (Fig. S1, symbols). The present work was carried out within the ViSA – IRP "**Vi**brations of **S**emiconductor **A**lloys – **I**nternational **R**esearch **P**artnership" project of "wall-less" international laboratory (2024 – 2027) funded by the "Lorraine University Excellence" program and supported by the "Excellence Initiative – Research University program at Nicolaus Copernicus University in Toruń (Poland)". The mesocenter of calculation EXPLOR at the Université de Lorraine (project 2019CPMXX0918) enabled the execution of *ab initio* calculations.

**Availability of data**
The data that support the findings of this study are available from the corresponding author upon reasonable request.

Supplementary material: Comment on *Infrared dielectric function of GaAs$_{1-x}$P$_x$ semiconductor alloys near the reststrahlen bands* [Appl. Phys. Lett. **123**, 172102 (2023)]

A. Elmahjoubi, T. Alhaddad, A.V. Postnikov and O. Pagès
LCP-A2MC, UR 4632, Université de Lorraine, 57000 Metz, France

In the main text, we outline an updated version of the PM for GaAs$_{1-x}$P$_x$ that explains, beyond existing CM and PM approaches, its phonon mode behavior, as apparent in the novel IR-ellipsometry $Im\{\varepsilon_r(\omega,x)\}$ data of Zollner *et al.*[1] (Fig. S1, symbols). Sec. I reports on the main features of the cluster model[3] (CM) and of the percolation model[4] (PM). A common terminology is introduced to facilitate the discussion. Both the TO-frequency and the TO-intensity aspects are covered. In Sec. II, an updated version of the PM for GaAs$_{1-x}$P$_x$ is presented (Fig. S2), supported by existing (Fig. S3) and novel (Fig. S4) *ab initio* phonon calculations at minimum (x~0,1 – TO-frequency aspect)[4] and maximum (x~0.5 – TO-intensity aspect) alloy disorder. This updated PM scheme for GaAs$_{1-x}$P$_x$ is directly confronted with Zollner's $Im\{\varepsilon_r(\omega,x)\}$ experimental data (Fig. S1, curves). An overview of the so far tested CM and PM approaches on GaAs$_{1-x}$P$_x$, including the current PM one, helps to evaluate their relative bias and merits. The overview focuses on the sensitive Ga-P phonon signal at the critical GaAs$_{0.725}$P$_{0.275}$ composition (Fig. S5), used as a case study. Last, for the sake of completeness, we show how the latter signal is impacted within the PM by deviating from the ideal random As↔P substitution (Fig. S6).

**SI. Main features of the cluster model (CM) and of the percolation model (PM) for GaAs$_{1-x}$P$_x$**

A critical comparison of the cluster model (CM) and of the percolation model (PM) used to explain the phonon spectra of GaAs$_{1-x}$P$_x$ has been conducted qualitatively in Ref. 4. Here, we recall the main features of the two models and set up a common jargon to facilitate their quantitative comparison. Basic features common to both models are first introduced, prior to presenting, in dedicated sections, the specificities of the two models concerning the $x$-dependence of the TO-frequencies and TO-intensities.

SI.A. CM and PM – common features

SI.A.1. Sensitivity of bond vibrations to the local environment

In their pioneering IR-reflectivity study of GaAs$_{1-x}$P$_x$ – contemporary with that of Verleur and Barker (VB),[3] Chen, Shockley and Pearson[2] grasped the GaAs$_{1-x}$P$_x$ phonon behavior in the simplest manner, via merely two {1×(Ga-As),1×(Ga-P)} oscillators, by voluntarily ignoring any fine structuring of the Ga-As and Ga-P reststrahlen bands. A successful variant of their random-element-isodisplacement (REI, 1-bond→1-phonon) model – worked out for the transverse optical (TO) modes – was subsequently designed as the modified-REI (MREI) model – applying also to the longitudinal optical (LO) modes,[S1] still a reference today.[S2] In the MREI model the like bonds of a given species (Ga-As or Ga-P) vibrate at the same (TO or LO) frequency at a given GaAs$_{1-x}$P$_x$ composition (1-bond→1-mode scheme), irrespectively of their local environment (varying from site to site in a disordered crystal). In brief, the MREI model is "blind" to the local alloy disorder, by construction.

Both the CM and the PM operate beyond the MREI model, in that they presume a sensitivity of bond vibrations to the local environment. One basic difference between the CM and the PM is that the local environment is defined in the real three-dimensional (3D) crystal for the CM and at one-dimension (1D) along the linear chain approximation for the PM.

SI.A.2. Scalar equations of motion

Both the CM and the PM consist of phenomenological models in which the equations of motion per atom (Ga, As, P in this case) are scalar, *i.e.*, written at one-dimension (1D) by describing the crystal along the linear chain approximation. This leads to consider only pure stretching forces between ions. The justification for such 1D-approaches is that optical vibrational techniques (such as far-IR spectroscopies) can only detect long-wavelength modes (due to the long wavelength of the light probe). In these modes, the vibrations of the like atoms of a given species replicate as such from unit cell to unit cell across the crystal. Such uniformity renders a 3D description superfluous; a mere 1D description suffices in principle. Only the optical modes do propagate at long wavelength. In $GaAs_{1-x}P_x$, these modes refer to vibrations of the rigid cation (Ga in this case) sublattice against the two rigid anion (As and P) sublattices (vibrating in phase but with different amplitudes), corresponding to effective Ga-As and Ga-P bond-stretchings.

SI.A.3. Focus on the TO modes

Both the TO and LO modes are accessible experimentally by IR-reflectivity and IR-ellipsometry, corresponding to atom vibrations perpendicular to the direction of propagation and parallel to it, respectively. Despite the ionicity of the chemical bonding in a zincblende crystal, the discussed TO modes do not carry any electric field.[S3] They assimilate with purely-mechanical harmonic oscillators (mass + spring-like restoring force). Such non-polar modes are interesting because they hardly couple and hence preserve the natural diversity of the phonon pattern of a complex system such as an alloy. This justifies our focus on the TO modes from now on. A similar focus was done by VB (see Fig. 7 of Ref. 3) and Cebulski *et al*. (see Fig. 3 of Ref. 6) in describing the phonon behavior of $GaAs_{1-x}P_x$. In contrast, the LO modes have both mechanic and electric character.[S2] The LO modes couple via their common electric field, resulting in dramatic distortions of the LO signal, even out of resonance conditions (an overview is given, *e.g.*, Fig. 9 of Ref. S4). This is prohibitive when searching to elucidate the nature of a multi-phonon pattern.

SI.A.4. Introduction of a common TO-terminology

In preamble to the CM vs. PM comparison, there is a question of terminology. Structurally, a zincblende semiconductor alloy such as $GaAs_{1-x}P_x$ consists of two intercalated cation (Ga) and anion (As,P) cubic face centered sublattices. As such, it can be decomposed at any $x$ value into a distribution of the same five elementary Ga-centered tetrahedron-cluster units with zero to four As atoms at the vertices (and P otherwise), being clear that the fraction of each cluster in $GaAs_{1-x}P_x$ is $x$-dependent. Basically, the TO mode due to the Ga-As or Ga-P bond-stretching (abbreviated $a$ and $p$, respectively) in a given Ga-centered $T$-tetrahedron (3D) unit with $n$ As atoms at the vertices (and P otherwise) forming the first-neighbor (superscript) environment of the involved Ga atom, is labelled $T^1_{n,(a,p)}$. For example, $T^1_{3,a}$ is the TO mode due to the Ga-As stretching (subscript $a$) in the first-neighbor (superscript 1) Ga-centered tetrahedron (main label $T$) unit with three (subscript 3) As atoms (and one P atom) at the vertices. The $T^1_{n,(a,p)}$-notation of a TO-oscillator in the CM directly transposes to $L^{1,2}_{n,(a,p)}$ in the PM by changing the main label so as to reflect the $L$-linear (1D) character of the TO-oscillator in this model. For example, $L^2_{3,a}$ is the TO mode due to the Ga-As stretching ($a$) in the 1D-cluster ($L$) forming the second-neighbor (2) shell of the vibrating bond (Ga-As), involving three (3) As atoms in total, *i.e.*, GaAs(Ga-As)GaAs in a developed 1D-notation.

SI.B. CM for $GaAs_{1-x}P_x$

In their CM, VB assume that, *e.g.*, the Ga-P bonds (of central interest) vibrate at (slightly) different frequencies in a given $GaAs_{1-x}P_x$ alloy depending on the $T^{(1)}_i$-cluster unit ($i$=0 to 4) they belong to, among four possible ones (with one to four P atoms at the vertices and As otherwise). The same applies to Ga-As vibrations. This generic 1-bond→4-mode TO scheme generates eight TO modes in total per $GaAs_{1-x}P_x$ alloy.

SI.B.1. TO frequencies.

In fitting their raw-IR reflectivity spectra of $GaAs_{1-x}P_x$ within their 1-bond→4-mode CM, VB assumed that the upper Ga-As submode and lower Ga-P submode originate from $T^1_4$ and $T^1_0$, respectively. No prior condition was imposed on the frequencies of the remaining submodes in each spectral range, left free in the fitting procedure. The best fitting was a regular sequence of the $T_n^{(1)}$- frequencies in the Ga-As spectral range, *i.e.*, $(T_{1,a}^1, T_{2,a}^1, T_{3,a}^1, T_{4,a}^1)$ – the frequency regularly increasing with the number of As atoms at the $T$-vertices,[3] together with an irregular cluster sequence for Ga-P, *i.e.*, $(T_{0,p}^1, T_{3,p}^1, T_{2,p}^1, T_{1,p}^1)$. Generally, all TO frequencies increase with $x$ (see Fig. 7 of Ref. 3). This conforms to intuition since by increasing the fraction $x$ of short Ga-P bonds in GaAs$_{1-x}$P$_x$ the lattice constant decreases, and hence the chemical bonds suffer a compression, giving rise to an upward shift of the TO frequencies.[S5]

Cebulski *et al.*[6] proposed an alternative version of the CM involving a regular Ga-P $(T_{3,p}, T_{2,p}, T_{1,p}, T_{0,p})$-sequence together with a regular Ga-As $(T_{1,a}^1, T_{2,a}^1, T_{3,a}^1, T_{4,a}^1)$-sequence. However, such re-arrangement would forcibly feed a counterintuitive assumption that the Ga-P TO frequencies tend to decrease with the P content $x$ (see Fig. 3 of Ref. 6) – at variance with the Ga-As ones.

Generally, no mechanism is proposed within the CM to justify the separation between the like TO frequencies due to a given bond depending on the $T_i^{(1)}$-clusters they belong to. This gives flexibility to arrange the four available TO frequencies per bond. In fact, the arrangement varies very much between the CM realizations of VB[3] and of Cebulski *et al.*[6]

SI.B.2. TO intensities.

In contrast, the TO-intensity aspect is fully formalized within the CM, as summarized below.

The strength of, *e.g.*, the GaP-like $T_{n,p}^1$ mode at a given GaAs$_{1-x}$P$_x$ composition is directly governed by the amount of oscillator strength awarded to this mode. This scales linearly with the corresponding fraction of oscillator in the crystal,[S1] *i.e.*, with the fraction $x_{n,p}^1$ of Ga-P bonds (symbolized $p$) involved in the $T_n^1$ clusters at the $x$-composition considered, given by

$$x_{n,p}^1 = \frac{(4-n)}{4} \times f_n^1(x), \tag{1a}$$

where $f_n^1(x)$ is the fraction of $T_n^{1)}$ clusters forming the GaAs$_{1-x}$P$_x$ crystal. A similar expression $x_{n,a}^1$ is derived for the Ga-As bonds (symbolized $a$), *i.e.*,

$$x_{n,a}^1 = \frac{n}{4} \times f_n^1(x). \tag{1b}$$

In the ideal case of a random As↔P substitution, $f_n^1(x)$ follows the binomial Bernoulli law, *i.e.*,

$$f_n^1(x) = \binom{4}{n}(1-x)^n \times x^{4-n}. \tag{2}$$

As for the phonon oscillator strength of the pure GaAs or GaP compound, defined as the jump in dielectric constant on crossing the phonon resonance, it expresses as follows[6] by force of the LST relation,[S3]

$$S_{a,p} = \varepsilon_{S,(a,p)} - \varepsilon_{\infty,(a,p)} = \varepsilon_{\infty,(a,p)} \times \frac{\omega_{L,(a,p)}^2 - \omega_{T,(a,p)}^2}{\omega_{T,(a,p)}^2}. \tag{3}$$

In this expression $\omega_{T,(a,p)}$ and $\omega_{L,(a,p)}$ are TO and LO Raman frequencies while $\varepsilon_{S,(a,p)}$ and $\varepsilon_{\infty,(a,p)}$ represent the low- (at $\omega \ll \omega_{T,(a,p)}$) and high-frequency (at $\omega \gg \omega_{T,(a,p)}$) dielectric constants of GaAs ($a$) and GaP ($p$).

In fact, VB could not achieve a fair modeling of their GaAs$_{1-x}$P$_x$ IR-reflectivity spectra unless assuming a pronounced deviation from the ideally random As↔P substitution towards clustering. This was formalized by introducing an order parameter $\beta$ (positive) in the probability $P_{XX}$ for a given site next to a substituent site occupied by $X$ (As or P) to be also occupied by $X$, according to,

$$P_{pp} = x + \beta \times (1-x) \tag{4a}$$
$$P_{aa} = (1-x) + \beta \times x. \tag{4b}$$

At $\beta=0$, $P_{XX} = P_X$, where $P_X$ is the probability of a given site to be occupied by $X$, corresponding to the $X$-fraction in the crystal. This situation refers to the random As↔P substitution. When $\beta$ is positive (the considered case by VB), $P_{XX} > P_X$, manifesting a trend towards clustering. The maximum $\beta$-value is 1. In this full-clustering case, $P_{XX}=1$, meaning that As and P are completely separated. Note that the trend towards clustering for a given substituent increases with the fraction of the other substituent in the crystal.

The $f_n^1(x, \beta)$ terms derived by VB[3] on the above basis are

$$f_0^1(x,\beta) = x \times [P_{pp} - x \times (1 - P_{pp}^2)], \tag{5a}$$
$$f_1^1(x,\beta) = 4 \times x^2 \times P_{pp} \times (1 - P_{pp}), \tag{5b}$$
$$f_2^1(x,\beta) = 6 \times (1-x)^2 \times (1 - P_{aa})^2, \tag{5c}$$
$$f_3^1(x,\beta) = 4 \times (1-x)^2 \times P_{aa} \times (1 - P_{aa}), \tag{5d}$$
$$f_4^1(x,\beta) = (1-x) \times [P_{aa} - (1-x) \times (1 - P_{aa}^2)], \tag{5d}$$

satisfying a basic sum rule for the total number of $T_i^1$-clusters in the GaAs$_{1-x}$P$_x$ crystal, i.e.,

$$\sum_{n=0}^{4} f_n^1(x,\beta) = 1. \tag{6}$$

The $x_{n,(a,p)}^1(x,\beta)$ terms satisfy similar sum rules related to the As and P fractions in the crystal, i.e.,

$$\sum_{n=1}^{4} x_{n,a}^1(x,\beta) = 1 - x, \tag{7a}$$
$$\sum_{n=0}^{3} x_{n,p}^1(x,\beta) = x. \tag{7b}$$

$\beta$, that governs the TO intensities via the $x_{n,p}^1$- terms (given by Eq. 1), was used as a free parameter by VB in their CM.[3] A unique $\beta$ value was adjusted across the GaAs$_{1-x}$P$_x$ composition domain, taken as characteristic of this alloy.

SI.A.3. Relative dielectric function.

The relative dielectric function of GaAs$_{1-x}$P$_x$ in its $(x,\beta)$-dependence is expressed in classical form used by VB – refer to Eq. (A11) in Ref. 3, i.e.,

$$\varepsilon_r(\omega,x,\beta) = \varepsilon_\infty(x) + \sum_{n=1}^{4} x_{n,a}^1(x,\beta) \times S_a \times \frac{\omega_{T,a}^2}{\omega_{Tn,a}^2(x,\beta) - \omega^2 - j \times \gamma_{n,a} \times \omega} + \sum_{n=0}^{3} x_{n,p}^1(x,\beta) \times S_p \times \frac{\omega_{T,p}^2}{\omega_{Tn,p}^2(x,\beta) - \omega^2 - j \times \gamma_{n,p} \times \omega}. \tag{8}$$

In this expression, $\omega_{T,(a,b)}$ are defined above, $j^2 = -1$, $\varepsilon_\infty(x)$ represents the electronic contribution (varying linearly with $x$) besides the phonon contribution summed over eight oscillators in total. Out of these latter, four relate to Ga-As (first sum) and the remaining four relate to Ga-P (second sum). Each oscillator is described by a Lorentzian function, with a resonance centered at the relevant $\omega_{Tn,(a,p)}(x,\beta)$ value of the TO frequency for Ga-As ($a$) or Ga-P ($p$) in the $T_n^1$ cluster at composition $x$ for the considered $\beta$ value. A damping term $\gamma_{n,(a,p)}$ (introduced via a friction force in the equations of motion per Ga-As or Ga-P oscillator) represents the full width at half maximum of the Lorentzian peak.

Note that VB alternatively express the damping term (refer to Eq. A11 in Ref. 3) as

$$\gamma_{n,(a,p)} = \Gamma_{n,(a,p)} \times \omega_{Tn,(a,p)}(x,\beta), \tag{9}$$

each constitutive term being adjusted from contour modeling of the raw IR-reflectivity spectra.

The $Im\{\varepsilon_r(\omega,x)\}$ spectra of GaAs$_{1-x}$P$_x$ are directly calculated from $\varepsilon_r(\omega,x,\beta)$ using Eq. (8) in this work (see below). Alternatively, Cebulski et al.[6] derived analytically $Im\{\varepsilon_r(\omega,x)\}$ from Eq. (8) and computed the resulting sum over multiple phonon-resonance terms, individually expressed as

$$\frac{S'_{n,(a,p)} \times \gamma_{n,(a,p)} \times \omega}{\left(\omega_{Tn,(a,p)}^2(x) - \omega^2\right)^2 + \gamma_{n,(a,p)}^2 \times \omega^2}, \tag{10}$$

with $S'_{n,(a,p)} = x_{n,(a,p)}^1 \times S_{(a,p)} \times \omega_{T(a,p)}^2$. \tag{11}

The theoretical IR-reflectivity at normal incidence/detection on GaAs$_{1-x}$P$_x$ is obtained from $\varepsilon_r(\omega,x,\beta)$ using the classical formula,[3]

$$R_{(\omega,x,\beta)} = \left|\frac{\sqrt{\varepsilon_r(\omega,x,\beta)} - 1}{\sqrt{\varepsilon_r(\omega,x,\beta)} + 1}\right|^2. \tag{12}$$

SI.C. PM for GaAs$_{1-x}$P$_x$

The PM offers a flexible – yet consistent – and comprehensible scheme to describe the phonon behavior of random GaAs$_{1-x}$P$_x$, in which the TO-frequencies and the TO-intensities are qualitatively and quantitatively formalized in their $x$-dependence, respectively, with *ab initio* calculations in support.

Out original first-neighbor version of the PM for GaAs$_{1-x}$P$_x$[4] supports two TO modes per bond, in principle, corresponding to the following series of four 1D-oscillators $\{P(Ga - As)Ga, As(Ga -$ $As)Ga$, $P(Ga-P)Ga$, $As(Ga-P)Ga$}. The related TO modes were marked {$TO_{Ga-As}^{P,1}$, $TO_{Ga-As}^{As,1}$, $TO_{Ga-P}^{P,1}$, $TO_{Ga-P}^{As,1}$} in Ref. 4. The subscript indicates the vibrating unit while the superscripts specify its local environment, *i.e.*, As or P (first superscript) in the first-neighbor shell (second superscript, added in this work). Using the $L$-notation introduced in this work (Sec. II.A.4) the same modes are labeled {$L_{1,a}^1$, $L_{2,a}^1$, $L_{0,p}^1$, $L_{1,p}^1$}.

Our second-neighbor version of the PM for GaAs$_{1-x}$P$_x$ designed in this work (see below) supports three TO oscillators per bond. For, say Ga-P, these are $GaP(Ga-P)GaP$, $GaAs(Ga-P)GaP$ and $GaAs(Ga-P)GaAs$. A basic rule tested on various Zn-based zincblende alloys,[10-12] states that the second-neighbor version of the PM distinguishes between like bond vibrations in totally "alien" environments and in the two remaining ones involving at least partly the "same" bond species. Applied to GaAs$_{1-x}$P$_x$, this rule comes to assimilate {$TO_{Ga-As}^{P,2}$, $TO_{Ga-As}^{As,2}$, $TO_{Ga-P}^{P,2}$, $TO_{Ga-P}^{As,2}$} with {$L_{1,a}^2$, $L_{2,a}^2 + L_{3,a}^2$, $L_{0,p}^2 + L_{1,p}^2$, $L_{2,p}^2$}. An overview is provided in Fig. S2, to facilitate the forecoming discussion.

SI.C.1. TO frequencies

Being lighter (by ~40%), Ga-P vibrates at a higher frequency (~360 cm$^{-1}$) than Ga-As (~260 cm$^{-1}$) in GaAs$_{1-x}$P$_x$.[1,3] In each spectral range, the PM predicts a bimodal fine structuring of the TO signal. The 1-bond→2-TO splitting is due to the local strain resulting from the contrast between the Ga-As and Ga-P bond lengths inherent to GaAs$_{1-x}$P$_x$-alloying. For example, a single As impurity substituting for P in GaP creates a local compressive strain because the Ga-As bond is longer[S6] than the Ga-P one (by ~3.5%). Hence, the few Ga-P bonds around As vibrate in "alien" (GaAs-like) environment at a higher frequency than the numerous matrix-like Ga-P bonds situated away from As, vibrating in "same" (GaP-like) environment. This results in a {dominant-$TO_{Ga-P}^P$, minor-$TO_{Ga-P}^{As}$} PM-type doublet at $x$~1, in order of increasing frequency. A single P impurity in GaAs likewise creates a local tensile strain, giving rise to a {minor-$TO_{Ga-As}^P$, dominant-$TO_{Ga-As}^{As}$} PM-type doublet for Ga-As at $x$~0, in this order of frequency.

The above Ga-As and Ga-P doublets predicted for GaAs$_{1-x}$P$_x$ within the PM based on the local strain are tested *ab initio* hereafter. For doing so, we extract additional information from *ab initio* (SIESTA code) phonon calculations previously done and discussed in Ref. 4. Such calculations are concerned with the phonon density of states projected at the center (corresponding to long wavelength) of the Brillouin zone (zone-center Ph-DOS) related to the matrix-like bonds of large (64-atom, 2×2×2) GaAs and GaP zincblende-type supercells containing a single impurity, noted GaAs:1P ($x$~0) and GaP:1As ($x$~1), respectively. The used version of the SIESTA code does not take into account the macroscopic electric field due to the ionicity of the chemical bonding in a zincblende crystal that is likely to be carried by a long wavelength optical mode. Hence the reported zone-center Ph-DOS directly assimilates with $Im\{\varepsilon_r(\omega,x)\}$ that captures the divergence $\varepsilon_r(\omega,x) \to \infty$ characteristic of the non-polar (deprived of electric field) transverse optic (TO) mode,[7,S3] as probed in a IR-reflectivity/ellipsometry experiment. Separate zone-center Ph-DOS are calculated per invariant-Ga atom from the first-neighbor shell of the isolated impurity (mode symbolized $iv$, shaded areas – Fig. S3) and per invariant-Ga atom situated away from the impurity (mode symbolized $iii$, curves). In the both Ga-As and Ga-P spectral ranges, the frequencies of the ($iii$) and ($iv$) modes arrange as predicted by considering the local strain (see above). The *ab initio* test is thus conclusive in the positive sense.

Entering detail, the Ga atoms from the first-neighbor shell of a single (As or P) impurity are involved in various local modes (minor shaded peaks) shifted away from the matrix-like zone-center Ph-DOS (curves), labelled A to E. We are only interested in those modes that represent opposite vibrations of the Ga and As/P ions, since otherwise the modes are not likely to be revealed by optical vibrational spectroscopies (Sec. SI.A.2). Hence, B, C and D, corresponding to in-phase displacements of the single P-impurity with its surrounding Ga atoms (mode B) and with rotation (mode C) or breathing (mode D) of the P atoms around or towards, respectively, the Ga atoms bridged to the single As-impurity (top views), are discarded in this sense. The only relevant features for our use are the A and E modes, corresponding to effective bond-stretching of the matrix-like bonds in the first-neighbor shell of the isolated P (mode A) and As (mode E) impurities. Within the PM terminology, A and E (shaded peaks) are naturally assigned in terms of the matrix-like bond-stretching in the "alien" environment ($iv$). In fact, both kinds of A and E vibration patterns (as sketched out) are involved in the two ($iv$)-type modes created by the single P and As impurities, depending on the four Ga atoms bridged to the latter impurities. As for the main peaks due to Ga situated beyond the first-neighbor shell of the P and As impurities (curves), they naturally refer to the matrix-like bond-stretching in "same" environment ($iii$) within the PM terminology. The $iii - iv$ frequency gaps for Ga-As and Ga-P are designated as $\Delta_{As}$ and $\Delta_P$ (double arrows in Fig. S3), respectively, to recollect with the used notation in Ref. 4. Note that the zone-center Ph-DOS remain basically unchanged by complexifying the impurity motif from a single impurity (upper Ph-DOS, Fig. S3) to a duo (lower Ph-DOS), or even a trio or a quatuor of impurities bridged by Ga (not shown), *i.e.*, by departing from the dilute-impurity limit.

Similar $\Delta_{As}$ and $\Delta_P$ frequency gaps were identified *ab initio* in Ref. 4 (using the same setup) by focusing on the impurity-duo. In fact, the impurity-duo is the minimal impurity motif that suffices to generate a separation between the in-chain ($ii$) mode (symbolized ↔ in Fig. S2) and the remaining out-of-chain ($i$) ones (symbolized ↮), naturally traceable to vibrations in "same" and "alien" environments within the PM, respectively. As for the single impurity, it necessarily vibrates in "alien" environment, and hence provides an impurity mode degenerated with ($i$) in principle (as apparent in Fig. 3 of Ref. 4).

Altogether, the above impurity- (full circles – Fig. S2, taken from Fig. 3 of Ref. 4) and matrix-related (hollow circles – this work) *ab initio* data support a "TO-frequency vs. $x$" plot for GaAs$_{1-x}$P$_x$ involving four quasi-parallel $\{TO^P_{Ga-As}, TO^{As}_{Ga-As}, TO^P_{Ga-P}, TO^{As}_{Ga-P}\}$ PM-branches – in order of frequency, as sketched out in Fig. S2. The numerical superscript (1, 2) specifying the (first, second)-neighbor shell of the vibrating unit is omitted at this stage because the current discussion is concerned with the TO frequencies only, while the superscript governs the TO intensities (see below).

The four GaAs$_{1-x}$P$_x$ branches, *i.e.*, two per bond species created by the local strain (as discussed above), are attached to eight *ab initio* TO frequencies at both ends of the composition domain, *i.e.*, four per bond species. We refer to the parent frequencies ($iv$) and to their minor satellites created by an isolated impurity ($iii$), at the one end of the composition domain, and to the frequencies of the out-of-chain ($ii$) and in-chain ($i$) modes due to the impurity-duo, at the other end, as discussed above. In a crude approximation the TO branches are taken straight, omitting the singularities in the Ga-P frequencies on crossing the bond percolation thresholds predicted in Ref. 4 (Fig. 2 therein). We recall that the latter critical compositions, achieved at $x$= 0.19 for Ga-P and at $x$=0.81 for Ga-As (in the random case), correspond to the first infinite connection of the minor bonds throughout the crystal, a purely statistical effect of the random As↔P substitution.[57] The justification for our omission is that the TO frequencies extracted from contour modeling of the raw IR-reflectivity spectra of GaAs$_{1-x}$P$_x$ by VB[3] are neither for nor against such singularities (see Fig. 2 of Ref. 4). Besides, the addressed compositions in the novel IR-ellipsometry data of Zollner *et al.*[1] strictly fall within the percolation regime (0.19≤ $x$ ≤0.81), and hence are not informative on what happens by crossing the bond percolation thresholds. As-pictured, the frequency gaps between two like phonon branches are comparable in the Ga-As and Ga-P spectral ranges, *e.g.*, ~15 cm$^{-1}$ at $x$=0.5, increasing in the parent-limit and decreasing in the impurity-limit.

The general orientation of branches is dictated by the macroscopic strain. The long (resp. short) Ga-As (resp. Ga-P) bonds are compressed (resp. tensed) on increasing the P (resp. As) fraction, resulting in an upward (resp. downward) shift of the related TO frequencies on departing from the parent values (Sec. SI.A.1). The four TO branches are thus upward-tilted in the TO-frequency vs. $x$ plot of Fig. S2.

SI.C.2. TO intensities

In Fig. S2, the intensities of the individual $\{TO_{Ga-As}^{P}, TO_{Ga-As}^{As}, TO_{Ga-P}^{P}, TO_{Ga-P}^{As}\}$ modes of GaAs$_{1-x}$P$_x$ are directly governed by the fractions $x_{n,(a,p)}^{(2)}$ – from now on the superscript in $x$-labelling is indicated within brackets so as to avoid any confusion with a square multiplication – of the underlying $L_{n,(a,p)}^2$ oscillators depending on the composition $x$.

For clarity, we refer back briefly to the original first-neighbor version of the PM for random ($\beta$=0) GaAs$_{1-x}$P$_x$ worked out in Ref. 4. The generic $\{TO_{Ga-As}^{P,1}, TO_{Ga-As}^{As,1}, TO_{Ga-P}^{P,1}, TO_{Ga-P}^{As,1}\}$ pattern – now equipped with a relevant numeral superscript – then identifies with $\{L_{1,a}^1, L_{2,a}^1, L_{0,p}^1, L_{1,p}^1\}$. The related oscillator fractions at $\beta$=0 are $x_{1,a}^1 = x \times (1-x)$, $x_{2,a}^1 = (1-x)^2$, $x_{0,p}^1 = x \times (1-x)$, $x_{1,p}^1 = x^2$. In Ref. 4, the two GaAs-like oscillators were merged into one (with oscillator fraction $x_a = 1-x$).

In the current second-neighbor version of the PM scheme for random ($\beta$=0) GaAs$_{1-x}$P$_x$ – sketched out in Fig. S2, the four-mode $\{TO_{Ga-As}^{P,2}, TO_{Ga-As}^{As,2}, TO_{Ga-P}^{P,2}, TO_{Ga-P}^{As,2}\}$ pattern now covers six elementary oscillators in total, i.e., $\{L_{1,a}^2, L_{2,a}^2, L_{3,a}^2, L_{0,p}^2, L_{1,p}^2, L_{2,p}^2\}$. Some regrouping is needed to achieve matching between the $TO$- and $L$-notations. One option – retained in Fig. S2 – is to distinguish between the like bonds of a given species depending on whether they vibrate in fully "alien" environments or in the remaining environments involving at least one bond of the "same" species, by analogy with earlier second-neighbor versions of the PM worked out for various Zn-based[10-12] semiconductor alloys. The resulting $L$-sequence is $\{L_{1,a}^2, L_{2,a}^2 + L_{3,a}^2, L_{0,p}^2 + L_{1,p}^2, L_{2,p}^2\}$. The same regrouping has been made, e.g., for Cd$_{1-x}$Zn$_x$Te (see Fig. 4 of Ref. 12). The related oscillator fractions in the random case ($\beta$=0) are $x_{1,a}^{(2)} = x^2 \times (1-x)$, $x_{(2,3),a}^{(2)} = (1-x)^3 + 2 \times x \times (1-x)^2$, $x_{(0,1),p}^{(2)} = x^3 + 2 \times x \times (1-x)^2$ and $x_{2,p}^{(2)} = x \times (1-x)^2$. The TO intensities scale accordingly, as represented by normalized blackened bar charts in Fig. S2 (arbitrarily oriented to the right). The bar charts offer a schematic, yet accurate, visual overview of the TO intensities across the entire GaAs$_{1-x}$P$_x$ composition domain.

In fact, Zollner's data[1] (Fig. S1, symbols) encourage an even more compact version of the PM in which all Ga-As oscillators are merged into one. This conforms to a basic rule in a zincblende alloy that, in practice, the PM-doublet generally shows up distinctly for the short bond only (Ga-P, in this case) and is hardly visible for the long one (Ga-As). The reason is that the short bond usually involves the small substituent. As such, the short bond has more room to distort than the long one to accommodate locally the contrast in bond length inherent to alloying, with concomitant effect on the TO frequencies, being generally more diversified for the short bond than for the long one. This results in a three-mode $\{TO_{Ga-As}, TO_{Ga-P}^{P,2}, TO_{Ga-P}^{As,2}\}$ pattern identifying with $\{L_{1,a}^2 + L_{2,a}^2 + L_{3,a}^2, L_{0,p}^2 + L_{1,p}^2, L_{2,p}^2\}$. The related fractions of the as-regrouped $L_{n,(a,p)}^2$-oscillators per TO mode, presently expressed in their $(x,\beta)$-dependence (established in detail in Ref. 11) – for the sake of completeness, are,

$$x_a(x,\beta) = 1-x, \tag{13a}$$

$$x_{(0,1),p}^{(2)}(x,\beta) = x \times \left(P_{pp} \times P_{ppp} + 2 \times (1-x) \times (1-P_{aa})\right), \tag{13b}$$

$$x_{2,p}^{(2)}(x,\beta) = (1-x)^2 \times (1-P_{aa}). \tag{13c}$$

In the central term, $P_{pp}$ is defined via $\beta$ along Eq. 4a – with exactly the same meaning, while $P_{ppp}$ is the probability to form a P-trio, expressed in a similar form as $P_{pp}$, i.e.,

$$P_{ppp} = x + \kappa \times x, \tag{14}$$

with $\kappa$ being related to $\beta$ via

$$\kappa = \beta \times P_{pp}^{-1}(x,\beta). \tag{15}$$

With this, the $x_p$-terms satisfy a basic sum rule, i.e.,

$$x_{(0,1),p}^{(2)}(x,\beta) + x_{2,p}^2(x,\beta) = x. \tag{16}$$

The relative dielectric function used in Ref. 4 to simulate the IR-reflectivity of GaAs$_{1-x}$P$_x$ taken by VB[3] within the original first-neighbor version of the PM operated at $\beta$=0 is

$$\varepsilon_r(\omega, x) = \varepsilon_\infty(x) + (1-x) \times S_a \times \frac{\omega_{Ta}^2}{\omega_{Ta}^2(x) - \omega^2 - j \times \gamma_a \times \omega}$$

$$+ \sum_{n=0}^{1} x_{n,p}^{(1)}(x) \times S_p \times \frac{\omega_{Tp}^2}{\omega_{Tn,p}^2(x) - \omega^2 - j \times \gamma_{n,p} \times \omega}. \tag{17}$$

Its $\beta$-dependent counterpart for the second-neighbor version of the PM worked out for random GaAs$_{1-x}$P$_x$ in this comment, used for contour modeling of the $Im\{\varepsilon_r(\omega, x)\}$ IR-ellipsometry data (Fig. S1, symbols) is given by,

$$\varepsilon_r(\omega, x, \beta) = \varepsilon_\infty(x) + (1-x) \times S_a \times \frac{\omega_{Ta}^2}{\omega_{Ta}^2(x) - \omega^2 - j \times \gamma_a \times \omega} + S_p \times \left( x_{(0,1),p}^{(2)}(x, \beta) \times \right.$$

$$\left. \frac{\omega_{Tp}^2}{\omega_{T(0,1),p}^2(x) - \omega^2 - j \times \gamma_{(0,1),p} \times \omega} + x_{2,p}^{(2)}(x, \beta) \times \frac{\omega_{Tp}^2}{\omega_{T2,p}^2(x) - \omega^2 - j \times \gamma_{2,p} \times \omega} \right). \tag{18}$$

A decisive test between the first- and second-neighbor versions of the PM for random GaAs$_{1-x}$P$_x$ is done *ab initio* (SIESTA code) by calculating the zone-center Ph-DOS on two large and fully relaxed (shape and atom positions) 4×4×4 primitive (rhombohedral, 128-atom) and 3×3×3 unit (cubic-face-centered, 216-atom) GaAs$_{1-x}$P$_x$ supercells ($x$=0.5) with special quasirandom ($\beta$=0) structure.[9] Different size and shape of supercells are considered to check the reproducibility of results. The two supercells are generated by using the Alloy Theoretic Automated Toolkit[S8] (ATAT) as the best randomized ones given the shapes and numbers of atoms of the supercells. Both consist of a zincblende-type alternance of cations and anions together with a random As↔P cation arrangement reproducing pair (14 Å) and multiple (triple – 14 Å and quadruple – 12 Å) correlations up to a large length scale (as specified in brackets). The used *ab initio* setup is the same as described in Ref. 4. Based on the $x_{n,(a,p)}^{(1,2)}$ fractions of $TO$-oscillator, the intensities of the two Ga-As or Ga-P TO submodes at $x$=0.5 should scale as ~1:1 in case of a sensitivity limited to first neighbors – a direct insight for GaAs$_{0.5}$P$_{0.5}$ is given in Fig. 4 of Ref. 4 (thick curves) – and as ~3:1 in case of a sensitivity extended up to second neighbors, with opposite (~3:1) asymmetries in the Ga-As and Ga-P spectral ranges in the latter case, as apparent in Fig. S2 (blackened bar charts, $x$=0.5).

The zone-center Ph-DOS of the two GaAs$_{0.5}$P$_{0.5}$ supercells are shown in Fig. S4. In both cases, the GaAs- and GaP-like Ph-DOS exhibit opposite asymmetries, towards low- and high-frequency, respectively, as predicted in case of a sensitivity of bond vibrations to second neighbors (refer to the blackened bar charts in Fig. S2). In fact, best (free) modeling of the distinct bimodal Ga-P signals generated by the 128-atom and 216-atom quasirandom supercells is achieved by using two Lorentzian functions (dotted curves, defined along Ref. S9) with similar phonon dampings and with intensity ratios scaling as ~2.4:1 and ~2.8:1, respectively, *i.e.*, close to ~3:1. The exact pairs of (TO frequencies, TO dampings, intensity) in cm$^{-1}$ for the 128-atom (352.4 – 372.3, 10.2 – 10.6, 2.05 – 0.87) and 216-atom (356.2 – 371.1, 11.6 – 11.0, 2.1 – 0.76) supercells are specified in brackets. This validates the second-neighbor scenario for GaAs$_{1-x}$P$_x$.

Summarizing, our current *ab initio* phonon calculations independently concerned with the TO frequencies (Sec. SI.B.1, $x$~0,1) and with the TO intensities (Sec. SI.B.2, $x$~0,5) support the updated ($\beta$=0)-version of the PM for random GaAs$_{1-x}$P$_x$ sketched out in Fig. S2. The update focuses on one main aspect of our original ($\beta$=0)-version of the PM,[4] *i.e.*, the sensitivity of bond vibrations earlier limited to first neighbors[4] is now extended over second neighbors.

**SII. Second-neighbor percolation model (PM) for random GaAs$_{1-x}$P$_x$**

This Sec. reports on various comparative tests on the second-neighbor version of the PM for random GaAs$_{1-x}$P$_x$ worked out in this comment. First, the latter version is tested in its $x$-dependence by referring to the novel and abundant $Im\{\varepsilon_r(\omega, x)\}$ data of Zollner *et al.*[1] (Fig. S1, symbols) – theory vs. experiment test. Second, an overview of the CM and PM approaches so far tested on GaAs$_{1-x}$P$_x$ helps to appreciate their relative merits and bias – theory vs. theory test. In doing so we focus on the sensitive Ga-P signal at the critical GaAs$_{0.725}$P$_{0.275}$ composition, corresponding to comparable intensities of the two Ga-P lines in Zollner's data.[1] Last, we test theoretically how the intensity interplay between the two lines forming the GaP-like PM-doublet of GaAs$_{1-x}$P$_x$ is impacted at the critical $x$~0.275 value by deviating from the ideally random As↔P substitution, for the sake of completeness.

SII.A. Second-neighbor PM scheme for GaAs$_{1-x}$P$_x$: theory vs. experiment

In main lines, the GaAs$_{1-x}$P$_x$ $Im\{\varepsilon_r(\omega,x)\}$ data derived by Zollner et al.[1] from their IR-ellipsometry data (Fig. S1, symbols) and reconstructed by Cebulski et al.[6] (Fig. 2 therein) from the raw IR-reflectivity spectra taken by VB,[3] reveal the same Ga-P bimodal pattern, hence robust and worth to discuss. The analogy is not so clear for Ga-As. A unique Ga-As feature is apparent in Zollner's data (Fig. S1) against two in those of Cebulski et al.,[6] less distinct than for Ga-P though – for reasons discussed in SI.C.2. Presumably, this is due to the difference in spectral resolutions used by VB[3] (2.5 cm$^{-1}$) and Zollner et al.[1] (4 cm$^{-1}$). Focusing on the best resolved Ga-As signal of Cebulski et al.,[6] it appears that the Ga-P and Ga-As bimodal signals echo each other, only that the $x$-dependent intensity interplay between the dominant and minor submodes is inverted for the two bonds. Hence, the critical compositions corresponding to intensity matching within each doublet are found at (nearly) symmetrical $x$ values for Ga-P (~0.275 – Fig. S1) and Ga-As (~0.85 – see Fig. 2 of Ref. 6). This cannot be merely fortuitous, suggesting that the (compact) Ga-As and (distinct) Ga-P fine structures have the same origin. In fact, symmetrical $x$-values for intensity matching within the GaAs- and GaP-like PM-doublets naturally emerge from the second-neighbor version of the PM for random GaAs$_{1-x}$P$_x$ outlined in Fig. S2 (refer to the blackened bar charts).

A more quantitative test is achieved by focusing on the Ga-P spectral range characterized by a distinct doublet in Zollner's data,[1] the Ga-As signal being apparently unimodal. Fair contour modeling of the $Im\{\varepsilon_r(\omega,x)\}$ GaAs$_{1-x}$P$_x$ spectra of Zollner et al.[1] is achieved (Fig. S1, curves) by using a simplified version of the generic PM scheme for random GaAs$_{1-x}$P$_x$ displayed in Fig. S2 in which the two Ga-As submodes are merged into one. This results in a three-mode $\{TO_{Ga-As}, TO_{Ga-P}^{P,2}, TO_{Ga-P}^{As,2}\}$ PM scheme for random GaAs$_{1-x}$P$_x$ identifying with $\{L_{1,a}^2 + L_{2,a}^2 + L_{3,a}^2, L_{0,p}^2 + L_{1,p}^2, L_{2,p}^2\}$. The related $\varepsilon_r(\omega,x)$ is given by Eq. (18). The TO intensities are directly governed by the explicit $x_{n,(a,p)}^{(2)}$-terms (Sec. B.2) acting as weighting factors on the parent oscillator strengths ($S_a$ and $S_p$). The latter are physical constants of the models calculated from Eq. (3) using the $\varepsilon_\infty$ values and the TO and LO frequencies given by Zollner et al.[1] (see Table S1 therein), i.e., (10.7, 268.2 cm$^{-1}$, 291.7 cm$^{-1}$) for GaAs and (8.9, 364.8 cm$^{-1}$, 403.3 cm$^{-1}$) for GaP. The TO frequencies and TO dampings are directly given by the peak positions and linewidths at half maximum apparent in the raw $Im\{\varepsilon_r(\omega,x)\}$ data (Fig. S1, symbols). Similar dampings (within few cm$^{-1}$) are used for the two Ga-P submodes, for simplicity. The used $\omega_{Tn,(a,p)}$ and $\gamma_{n,(a,p)}$ values depending on the composition $x$ are reported in Table S1. The $\omega_{Tn,(a,p)}(x)$ data sets (symbols in Fig. S2) are globally consistent with ab initio predictions (straight-plain lines) for what regards the Ga-P splitting. The experimental and theoretical values are not equal, though. This is due to a well-known bias of the linear density approximation used in our ab initio calculations to overestimate the bond force constants.[4] Generally, in Fig. S1, the overall agreement between experiment and theory is excellent given the absence of free parameter. This supports our longstanding view that the phonon mode behavior of GaAs$_{1-x}$P$_x$ basically follows the PM.[4]

SII.B. CM vs. PM comparative overview – GaAs$_{0.725}$P$_{0.275}$ as a case study

In this Sec., we recapitulate and compare the successive CM and PM approaches so far used to model the GaAs$_{1-x}$P$_x$ phonon spectra, i.e., four in fact – including the current one. We focus on $Im\{\varepsilon_r(\omega,x)\}$ that directly informs on the TO modes – of most interest in this work (as explained at the beginning of Sec. I) – and specifically address the sensitive Ga-P signal characterized by a distinct bimodal pattern at any $x$ value in Zollner's data (Fig. S1, symbols) – at variance with the Ga-As signal (unimodal in Zollner's data). As a case study, the discussion is placed at $x$=0.275, corresponding to comparable intensities of the two Ga-P lines. For the sake of consistency, all models are discussed in reference to the same observable, i.e., the $Im\{\varepsilon_r(\omega, x = 0.275)\}$ Ga-P spectrum of Zollner et al.[1] (Fig.

S5, symbols). This composition is moreover relevant since it is common (within half percent) to VB-Cebulski's data[3,6] ($x$=0.28) and to Zollner's data[1] ($x$=0.275). We emphasize that the various theoretical CM-curves reported in Fig. S5 are reconstructed by using the strict sets of ($\omega_{Tn,(a,p)}$, $x_{n,(a,p)}$, $\gamma_{n,(a,p)}$, $S_a$, $S_b$) input parameter terms – listed in Table S2, as adjusted by VB (Ref. 3) and Cebulski et al. (Ref. 6). To facilitate the comparison between the first- (Ref. 4) and second-neighbor (this work) versions of the theoretical PM-curves, the same set of input parameters are used in both cases (listed in Table S2).

The first series of reported theoretical data (Fig. S5a) refers to direct modeling of the raw IR-reflectivity spectra of GaAs$_{1-x}$P$_x$ taken by VB.[3] This includes the original version of the CM worked out by VB,[3] and also our own original first-neighbor version of the PM (Ref. 4). The theoretical IR-reflectivity given by Eq. (12) was adjusted to experiment using a classical form of $\varepsilon_r(\omega, x)$, given by Eq. (8) for the CM and by Eq. (18) for the PM. In each case, $Im\{\varepsilon_r(\omega, x)\}$ is reconstructed as a by-product of the fitting procedure (plain curves, Fig. S4a) offering an access to all constitutive ($\omega_{Tn,(a,p)}$, $x_{n,(a,p)}$, $\gamma_{n,(a,p)}$) parameters of $\varepsilon_r(\omega, x)$. The second series of theoretical data (plain curves Fig. S5b) refers to direct $Im\{\varepsilon_r(\omega, x)\}$ adjustments. Cebulski et al.[6] used their revised version of the CM considering a generic phonon contribution to $Im\{\varepsilon_r(\omega, x)\}$ given by Eq. (10). We operated the updated (second-neighbor) version of the PM worked out in this comment using Eq. (18).

The parent oscillator strengths ($S_a$ and $S_a$) are physical constants of all models, calculated from Eq. (3) via $\varepsilon_s - \varepsilon_\infty$ (Ref. 6 – Table 1) or using the ($i$) $\varepsilon_\infty$ values, ($ii$) TO frequencies and ($iii$) LO frequencies. The latter three values measured by Zollner et al.[1] (Table SI), used in this work, are (10.7, 268.2 cm$^{-1}$, 291.7 cm$^{-1}$) for GaAs and (8.9, 364.8 cm$^{-1}$, 403.3 cm$^{-1}$) GaP. The values used by Pagès et al. in Ref. 4 are (11, 269 cm$^{-1}$, 292 cm$^{-1}$) for GaAs and (9.1, 365 cm$^{-1}$, 405 cm$^{-1}$) GaP. This ends up in slightly different (GaAs, GaP) phonon oscillator strengths between authors, i.e., (2.07, 1.73) for VB (directly read from Fig. 8 of Ref. 3), (1.96, 2.10) for Pagès et al.,[4] (2.07, 2.0) for Cebulski et al. (table 1 of Ref. 6) and (1.96, 1.98) for Zollner et al.[1]

Below, we discuss the CM and the PM separately, for clarity, starting with the CM.

Both VB[3] and Cebulski et al.[6] implemented the CM beyond its basic ($\beta$=0) version for GaAs$_{1-x}$P$_x$, at different levels of sophistication. The two versions are presented with (Fig. S4, central panels) and without (upper panels) the related sophistications, to better appreciate their impact. In fact, as apparent in Fig. S4 (upper panels), the two raw versions of the CM for random ($\beta$=0) GaAs$_{1-x}$P$_x$ (central panels) fail to mimic the experimental bimodal Ga-P pattern (symbols). Either the lower Ga-P line (VB) or the upper Ga-P line (Cebulski et al.[6]) is missing. Note that in both cases the missing line relates to $T_0$. In fact, the $T_{0,p}^1$ line and the ($T_{3,a}^1$, $T_{2,p}^1$, $T_{1,p}^1$) series are swapped in the two descriptions, as already mentioned (Sec. I.A.1). From there, two deviations from the basic ($\beta$=0)-CM scheme were envisaged to retrieve the experimental balance between the two Ga-P lines.

VB[3] considered a pronounced trend towards local clustering, corresponding to a large positive $\beta$ value (i.e., 0.75). Structurally – referring to Eq. (5), this favors an over-representation of the "homo" $T_0$ (otherwise hardly represented in GaAs$_{0.725}$P$_{0.275}$ in case of a random As↔P substitution at such small P content) and $T_4$ clusters in the crystal (with P and As only at the vertices, respectively) at the expense of the remaining "hetero" ($T_3$, $T_2$, $T_1$) ones (with both As and P atoms at the vertices). The net effect on the GaP-like $Im\{\varepsilon_r(\omega, x = 0.275)\}$ spectra is to channel the available Ga-P oscillator strength from the high-frequency ($T_{3,p}^1$, $T_{2,p}^1$, $T_{1,p}^1$) lines towards the low-frequency $T_{0,p}^1$ line (compare the upper and central panels of Fig. S4a), thus improving the intensity balance between the two Ga-P lines.

Cebulski et al.[6] alternatively assumed a random As↔P substitution ($\beta$=0) but introduced an additional phonon mode (APM – central panel of Fig. S4b) to fill the void left by the missing Ga-P line (Fig. S4b, upper panel) besides the four ($T_{3,a}^1$, $T_{2,p}^1$, $T_{1,p}^1$, $T_{0,p}^1$) canonical (GaP-like) phonon modes (CPM) provided by the CM. However, the APM is not assigned. Cebulski et al.[6] argue that the APM is not a nominal GaP-like TO mode. Otherwise, this would challenge a basic sum rule concerned with the linear $x$-variation of the GaP-like oscillator strength for the TO modes.[S1]

The above two CM approaches pose problems. The far-from-random As↔P substitution in GaAs$_{1-x}$P$_x$ presumed by VB [3] was invalidated by subsequent extended X-ray absorption fine structure measurements.[5] The APM of Cebulski *et al.*[6] is not assigned. Besides, in either case there is no solid prediction concerning the arrangement of the TO frequencies. In fact, the ($T^1_{3,p}$, $T^1_{2,p}$, $T^1_{1,p}$) and $T^1_{0,p}$ lines are permuted in the two approaches.

Such problems are avoided with the PM. Both reported PM approaches (Fig. S5, lower panels) operate at a random As↔P substitution ($\beta$=0) and involve regular GaAs$_{1-x}$P$_x$ TO modes. Further, the PM is predictive regarding both the TO intensities (exactly) and the TO frequencies (in main lines), with *ab initio* calculations in support (Sec. S.I.B). Our original first-neighbor version of the PM (tested on the raw IR-reflectivity spectra of GaAs$_{1-x}$P$_x$ taken by VB)[4] fails to achieve the exact Ga-P intensity balance observed experimentally at $x$=0.275 (Fig. S5a, lower panel). The experimental balance is well replicated by using the current/updated second-neighbor version of the PM (Fig. S5b, lower panel).

From the overview given in Fig. S5, it emerges that the second-neighbor version of the PM for random GaAs$_{1-x}$P$_x$ naturally (*i.e.*, without any bias) explains its phonon mode behavior, beyond existing approaches.

Last, for the sake of completeness, we briefly test the second-neighbor version of the PM for GaAs$_{1-x}$P$_x$ in comparison with experiment – referring to Zollner's $Im\{\varepsilon_r(\omega,x)\}$ data – with respect to $\beta$=0. As already mentioned, $\beta$ informs on the nature of the atom substitution, as to whether this is ideally random ($\beta$=0) or due to clustering ($\beta$>0) or anticlustering ($\beta$<0). The test is again done at $x$=0.275, for the sake of consistency with Fig. S5. In the reported theoretical $Im\{\varepsilon_r(\omega,x)\}$ data (Fig. S6), even a slight $\beta$-deviation towards clustering ($\beta$=0.15) or anticlustering ($\beta$=-0.15) – using the $\beta$-dependent versions of the $x^{(2)}_{n,p}$-terms given by Eq. (13) – suffices to sweep away from the exact Ga-P intensity balance (achieved at $\beta\sim0$), creating an appreciable deviation (seen by eye) between experiment (symbol) and theory (plain lines). In the clustering case, the sense of the distortion predicted within the PM is consistent with that found *ab initio* for the ZnSe-like TO signal of Cd$_{1-x}$Zn$_x$Se (refer to the curved arrow in Fig. 10 of Ref. 11) that exhibits the same three-mode $\{TO_{Cd-Se}, TO^{Zn,2}_{Zn-Se}, TO^{Cd,2}_{Zn-S}\}$ second-neighbor PM scheme as GaAs$_{1-x}$P$_x$.

More generally, $\beta$=0 is a solid estimate for contour modeling of $Im\{\varepsilon_r(\omega,x)\}$ GaAs$_{1-x}$P$_x$ data across the composition domain, as apparent in Fig. S1. This is consistent with extended-x-ray-absorption fine structure (EXAFS) measurements of Wu *et al.*[5] that revealed GaAs$_{1-x}$P$_x$ as a random alloy.

**SUPPLEMENTARY MATERIAL ONLY – REFERENCES**

**Table S1:** Input parameters (defined in text) used to calculate the $Im\{\varepsilon_r(\omega,x)\}$ spectra of GaAs$_{1-x}$P$_x$ displayed in Fig. S1.

| $GaAs_{1-x}P_x$ | Mode $^2_{n,p}$ | $\omega_{T_{n,p}}$ $(cm^{-1})$ | $\gamma_{n,p}$ $(cm^{-1})$ | $x^{(2)}_{n,p}$ |
|---|---|---|---|---|
| $x = 0$ | | 269 | 3.5 | 0 |
| $x = 0.205$ | $L^2_{(0,1),p}$ | 348 | 8.2 | 0.0754 |
| | $L^2_{2,p}$ | 357.2 | 10.5 | 0.1296 |
| $x = 0.232$ | $L^2_{(0,1),p}$ | 348.2 | 8.1 | 0.0952 |
| | $L^2_{2,p}$ | 357.2 | 11 | 0.1368 |
| $x = 0.240$ | $L^2_{(0,1),p}$ | 347.8 | 8.4 | 0.1014 |
| | $L^2_{2,p}$ | 357.5 | 11.4 | 0.1386 |
| $x = 0.275$ | $L^2_{(0,1),p}$ | 347.5 | 9.5 | 0.1305 |
| | $L^2_{2,p}$ | 359.5 | 12.5 | 0.1445 |
| $x = 0.435$ | $L^2_{(0,1),p}$ | 351 | 10.1 | 0.2961 |
| | $L^2_{2,p}$ | 362 | 11.5 | 0.1389 |
| $x = 0.469$ | $L^2_{(0,1),p}$ | 352.2 | 12.3 | 0.3368 |
| | $L^2_{2,p}$ | 363.4 | 12.3 | 0.1322 |
| $x = 0.475$ | $L^2_{(0,1),p}$ | 352.95 | 9.9 | 0.3441 |
| | $L^2_{2,p}$ | 364.8 | 10.6 | 0.1309 |
| $x = 0.615$ | $L^2_{(0,1),p}$ | 355 | 7.1 | 0.5238 |
| | $L^2_{2,p}$ | 369.5 | 8.9 | 0.0912 |
| $x = 0.710$ | $L^2_{(0,1),p}$ | 357 | 9.9 | 0.6503 |
| | $L^2_{2,p}$ | 372.2 | 8.9 | 0.0597 |
| $x = 0.810$ | $L^2_{(0,1),p}$ | 361.2 | 7.2 | 0.7808 |
| | $L^2_{2,p}$ | 378 | 9.8 | 0.0292 |
| $x = 1$ | | 364.2 | 4.7 | 1 |

**Table S2:** Input parameters (defined in text) used to calculate the individual GaP-like CM- and PM-type $Im\{\varepsilon_r(\omega, x)\}$ $L_{n,p}^{(1,2)}$ lines for GaAs$_{1-x}$P$_x$ at $x$=0.275 displayed in Fig. S5.

| Model | Mode $_{n,p}^{1,2}$ | $\omega_{T_{n,p}}$ $(cm^{-1})$ | $\gamma_{n,p}$ $(cm^{-1})$ | $x_{n,p}^{1,2}$ |
|---|---|---|---|---|
| PM $1^{st}$ neighbors [4] | $L_{0,p}^1$ | 347.5[a] | 9.5[a] | 0.078 |
|  | $L_{1,p}^1$ | 359.5[a] | 12.5[a] | 0.202 |
| PM $2^{nd}$ neighbors [this work] | $L_{0,p}^2$ | 347.5 | 9.5 | 0.022 |
|  | $L_{1,p}^2$ | 347.5 | 9.5 | 0.113 |
|  | $L_{2,p}^2$ | 359.5 | 12.5 | 0.145 |
| CM ($\beta = 0.75$) [3] | $T_{0,p}^1$ | 346.4 | 10.4[b] | 0.204 |
|  | $T_{1,p}^1$ | 364.4 | 6.9[b] | 0.035 |
|  | $T_{2,p}^1$ | 361.5 | 6.5[b] | 0.008 |
|  | $T_{3,p}^1$ | 358.4 | 5.7[b] | 0.034 |
| CM ($\beta = 0$) [3] | $T_{0,p}^1$ | 343.85 | 10.3[b] | 0.006 |
|  | $T_{1,p}^1$ | 365.45 | 6.9[b] | 0.047 |
|  | $T_{2,p}^1$ | 362.65 | 6.5[b] | 0.122 |
|  | $T_{3,p}^1$ | 359.00 | 5.7[b] | 0.105 |
| Cebulski CM (CPM, $\beta = 0$) [6] | $T_{0,p}^1$ | 362.3 | 9.0 | 0.006 |
|  | $T_{1,p}^1$ | 358.0 | 12.1 | 0.047 |
|  | $T_{2,p}^1$ | 351.3 | 13.0 | 0.122 |
|  | $T_{3,p}^1$ | 345.6 | 11.0 | 0.105 |

| Model | Mode $_{n,p}^{1,2}$ | $\omega_{T_{n,p}}$ $(cm^{-1})$ | $\gamma_{n,p}$ $(cm^{-1})$ | $S'_{n,p}$ $(cm^{-2})$[c] |
|---|---|---|---|---|
| Cebulski CM (CPM + APM, $\beta = 0$) [6] | APM | 362.3 | 9.0 | 14000 |
|  | $T_{1,p}^1$ | 358.0 | 12.1 | 21000 |
|  | $T_{2,p}^1$ | 351.3 | 13.0 | 32500 |
|  | $T_{3,p}^1$ | 345.6 | 11.0 | 26500 |

[a] The same frequencies and damping are used for the first- and second-neighbors versions of the PM.
[b] $\gamma_{n,p} = \Gamma_{n,p} \, \omega_{T_{n,p}}$, Eq. (A11) [3]
[c] $S'_{n,p} = x_{n,p}^1 \, S_{n,p} \, \omega_{T_{n,p}}$

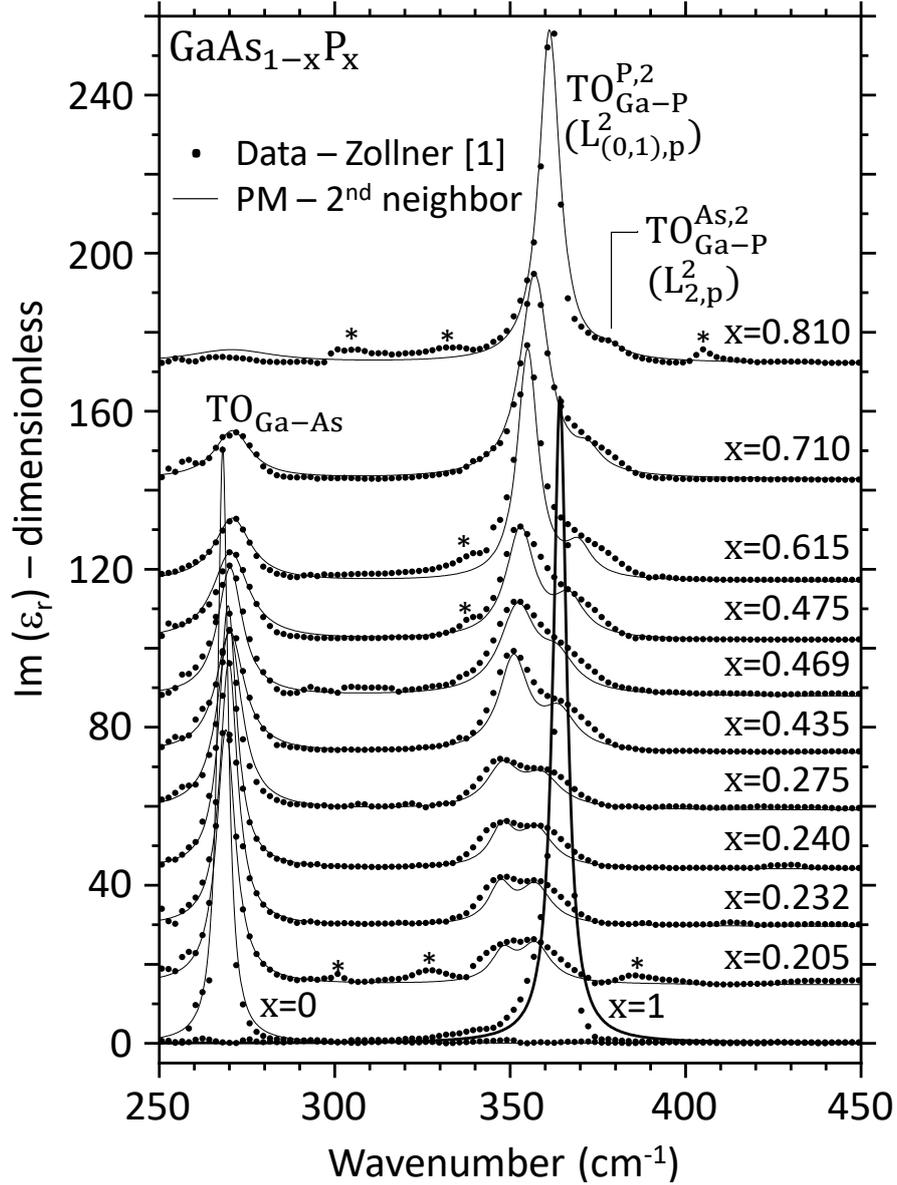

**Fig. S1: GaAs$_{1-x}$P$_x$ $Im\{\varepsilon_r(\omega,x)\}$ spectra.** Experimental (symbols, data provided by Zollner et al.[1]) and PM-based theoretical (lines) $Im\{\varepsilon_r(\omega,x)\}$ data. Most raw data are vertically shifted from zero for clarity. The theoretical curves are calculated using Eq. (18) within a simplified second-neighbor version of the PM scheme for random ($\beta$=0) is GaAs$_{1-x}$P$_x$ over the generic one sketched out in Fig. S2 (see text). The used input parameters are given in Table S1. The individual peaks are labeled by using both the crude/convenient TO-notation and the underlying/accurate $L$-notation, for clarity. The asterisks mark spurious features omitted in the model.

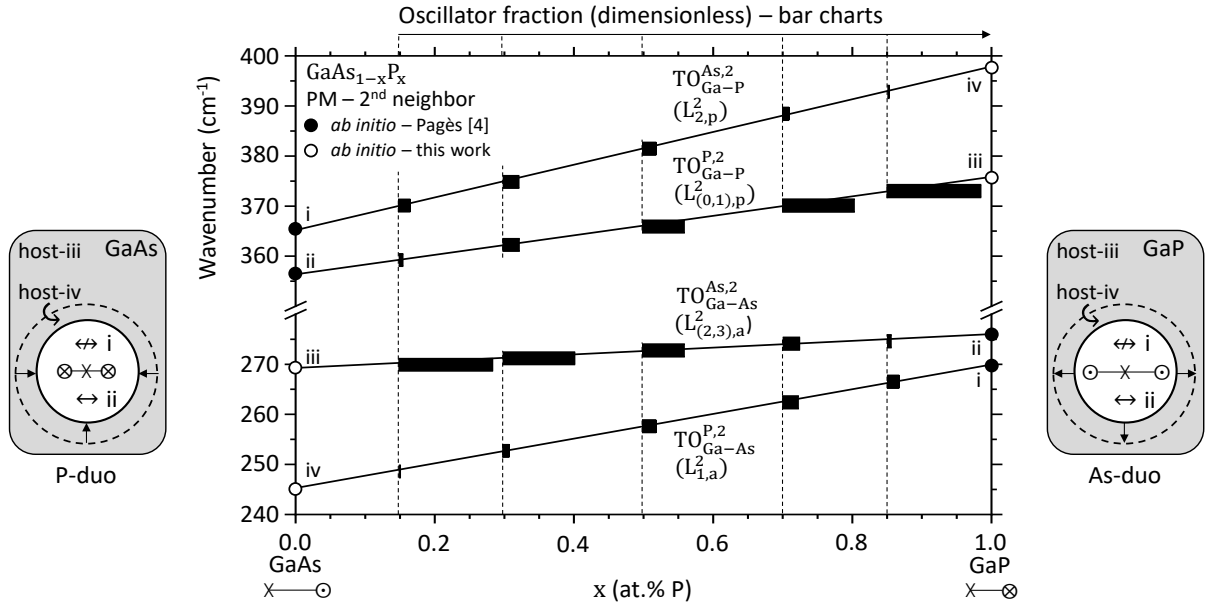

**Fig. S2: Generic second-neighbor PM scheme for random GaAs$_{1-x}$P$_x$.** Four-mode second-neighbor PM for the TO modes of random ($\beta$=0) GaAs$_{1-x}$P$_x$. The individual TO modes are labeled via the double TO-$L$ notation (see text), for clarity. The individual fractions of oscillators, derived via Eq. (13) using $\beta$=0 are represented by normalized blackened bar charts (arbitrarily oriented to the right), for a direct visual insight. Linear TO-frequency vs. $x$ variations are considered, attached to limit ($x\sim$0,1) values derived *ab initio* (symbols). The limit frequencies (labeled $i-iv$) are determined from a duo of impurity (As or P) bridged by Ga, as sketched out, isolated into a host matrix of the other type (GaP or GaAs, correspondingly). The inner-plain circle line defines the duo-motif. The PM distinguishes between the ($i$) out-of-chain and ($ii$) in-chain modes, symbolized ↔ and ↮, corresponding to vibrations of the impurity in "same" and "alien" environments, respectively. The area between the outer-dashed and inner-plain circle lines sets a host-matrix zone subject to the local tensile/compressive strain (as schematically represented by inward/outward arrows, correspondingly) created by the impurity-duo of short/long bonds (as sketched out beneath the abscissa axis). The host vibrations in "same" and "alien environments", *i.e.*, ($iii$) away from the duo-impurity motif and ($iv$) close to it, respectively, are accordingly distinguished (Fig. S3).

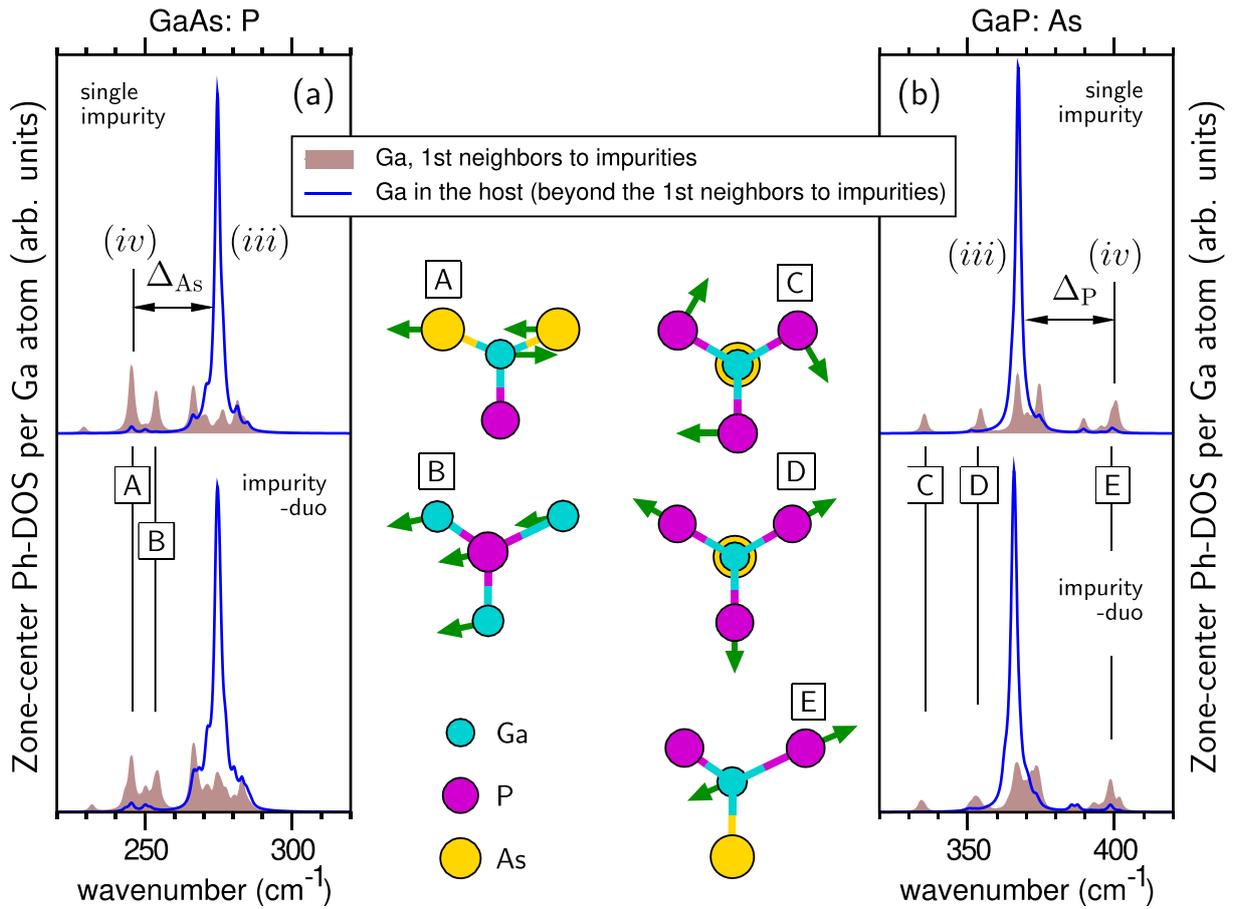

**Fig. S3: Zone center Ph-DOS of impurity-diluted GaAs$_{1-x}$P$_x$ (64-atom) supercells ($x$~0,1).** TO-like zone-center Ph-DOS per Ga atom generated *ab initio* (SIESTA code) at minimum alloy disorder ($x$=0,1) using large (64-atom) zincblende GaAs$_{1-x}$P$_x$ supercells containing either a unique impurity or a duo of impurity bridged by Ga. In each case, a distinction is made between Ga atoms situated within the first-neighbor shell of the impurity motif (shaded areas) and away from it (curves). For main peaks due to the first-neighbor shell, labelled A to E, the eigenvector structure of representative vibration modes is schematically shown. Out of these, the relevant modes for the PM are also labelled using the same *iii*/*iv* terminology as in Fig. S2 – for the sake of consistency. $\Delta_{As}$ and $\Delta_P$ mark the *iii-iv* frequency gaps within the Ga-As and Ga-P PM-doublets in the related parent limits.

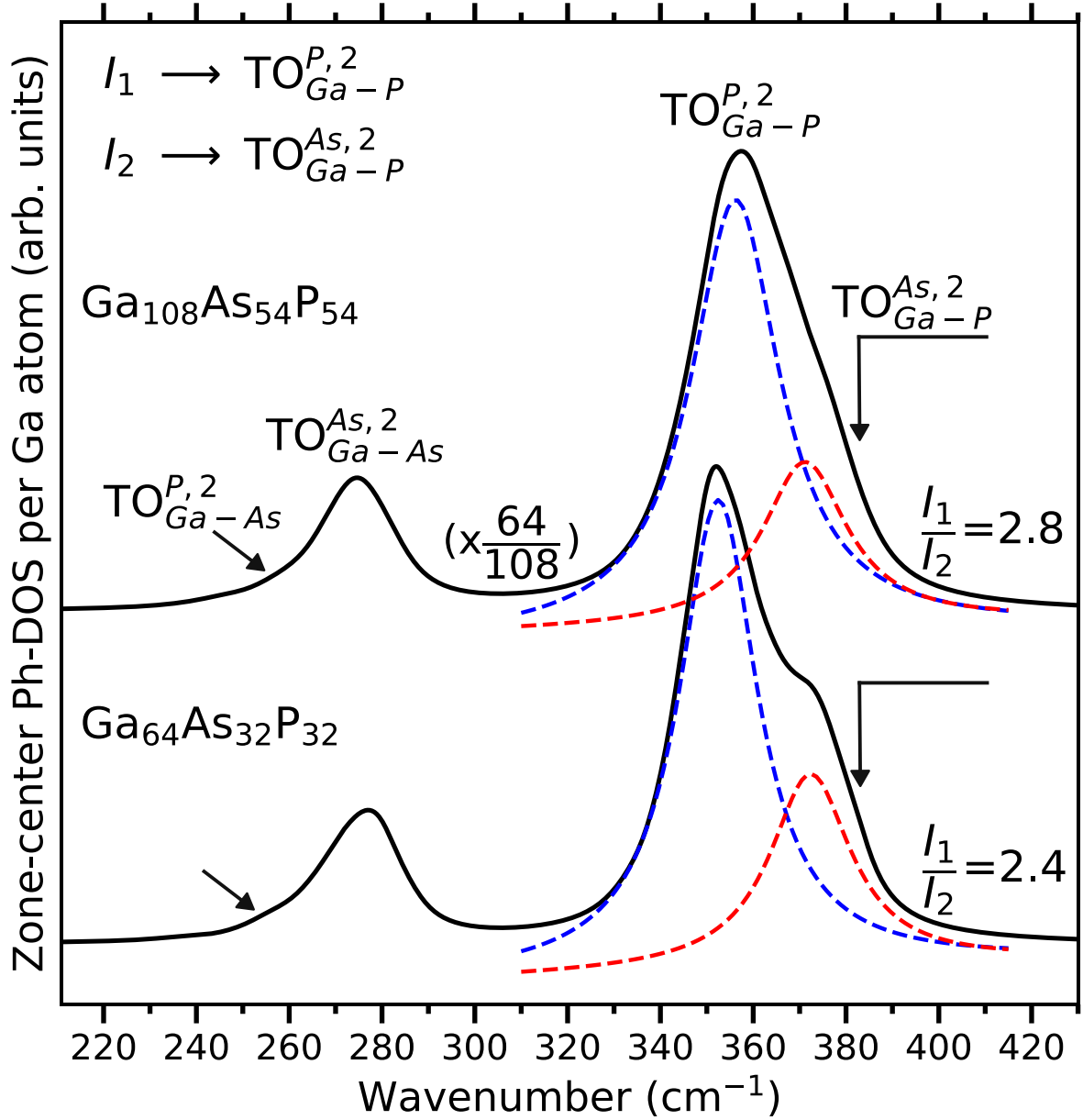

**Fig. S4: Zone-center Ph-DOS of special quasirandom GaAs$_{0.5}$P$_{0.5}$ supercells.** TO-like zone-center Ph-DOS per Ga atom generated *ab initio* (SIESTA code) using large 4×4×4-primitive (rhombohedral, 128-atom) and 3×3×3-unit (cubic-face-centered, 216-atom) zincblende-type GaAs$_{1-x}$P$_x$ supercells with special quasirandom structure at maximum alloy disorder ($x$=0.5). The two curves are normalized per Ga atom – as specified in brackets. The individual TO modes are labeled using the same crude/convenient TO-notation and underlying/accurate $L$-notation (see text) as in Fig. S2, to facilitate comparison of the TO intensities (scaling as the bar charts in Fig. S2). The GaP-like bimodal signal is freely adjusted by two Lorentzian functions (dotted curves – parameters are given in text) with intensity ratios scaling as specified ($I_1/I_2$).

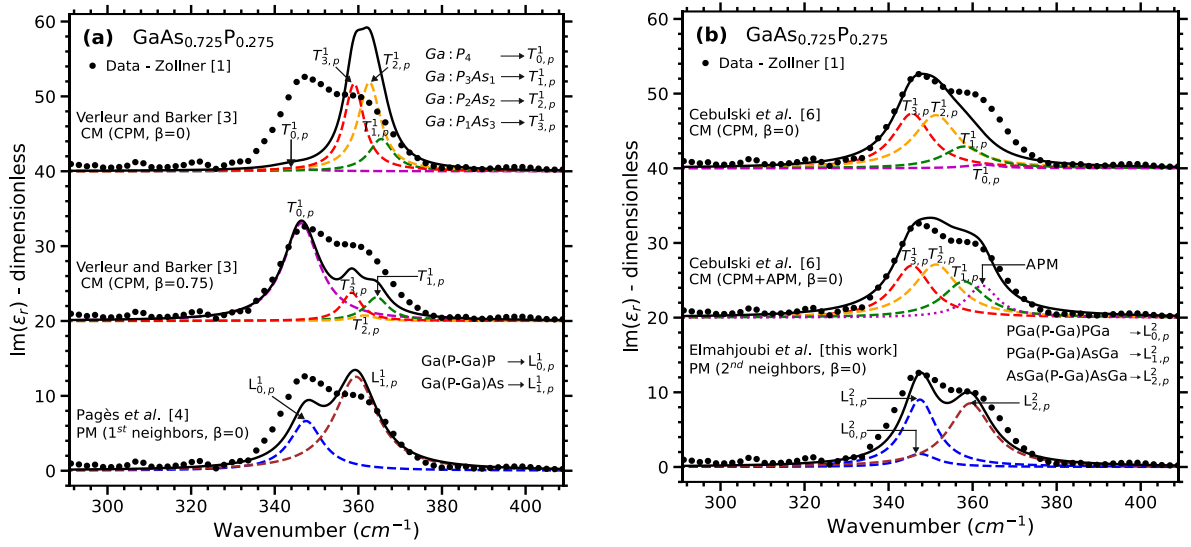

**Fig. S5: GaP-like $Im\{\varepsilon_r(\omega,x)\}$ spectra of GaAs$_{0.725}$P$_{0.275}$ – Experiment vs. theory.** Various theoretical $Im\{\varepsilon_r(\omega,x)\}$ GaP-like signals for GaAs$_{1-x}$P$_x$ at $x$=0.275 (plain curves) juxtaposed with corresponding experimental data obtained by Zollner et al.[1] (symbols). **(a)** Theoretical series obtained by contour modeling of the raw IR-reflectivity spectra taken by Verleur and Barker (VB).[3] The original CM scheme developed by VB is implemented in the random ($\beta$=0, upper panel) and non-random/clustered ($\beta$=0.75, central panel) cases, using Eq. (8). The original first-neighbor PM for random GaAs$_{1-x}$P$_x$ ($\beta$=0)[4] is implemented using Eq. (17). **(b)** Theoretical series obtained by contour modeling of $Im\{\varepsilon_r(\omega,x)\}$ as reconstructed from the raw IR-reflectivity[3] or IR-ellipsometry[1] GaAs$_{0.725}$P$_{0.275}$ spectra taken by VB[3] and Zollner et al.,[1] respectively. The variant of the CM scheme operated by Cebulski et al.[6] in the random case ($\beta$=0) is implemented without (upper panel) and with (central panel) an additional phonon mode (APM) besides the "canonical" phonon modes (CPM), using Eq. (10). The current second-neighbor variant of the PM for random ($\beta$=0, lower panel) GaAs$_{1-x}$P$_x$ is calculated by using Eq. (18). The individual TO modes are labeled along either the $T$-tetrahedron notation (CM) or the $L$-linear one (PM). The input parameters related to the individual TO lines (dotted curves) are given in Table S2. Colors are used to distinguish between individual modes.

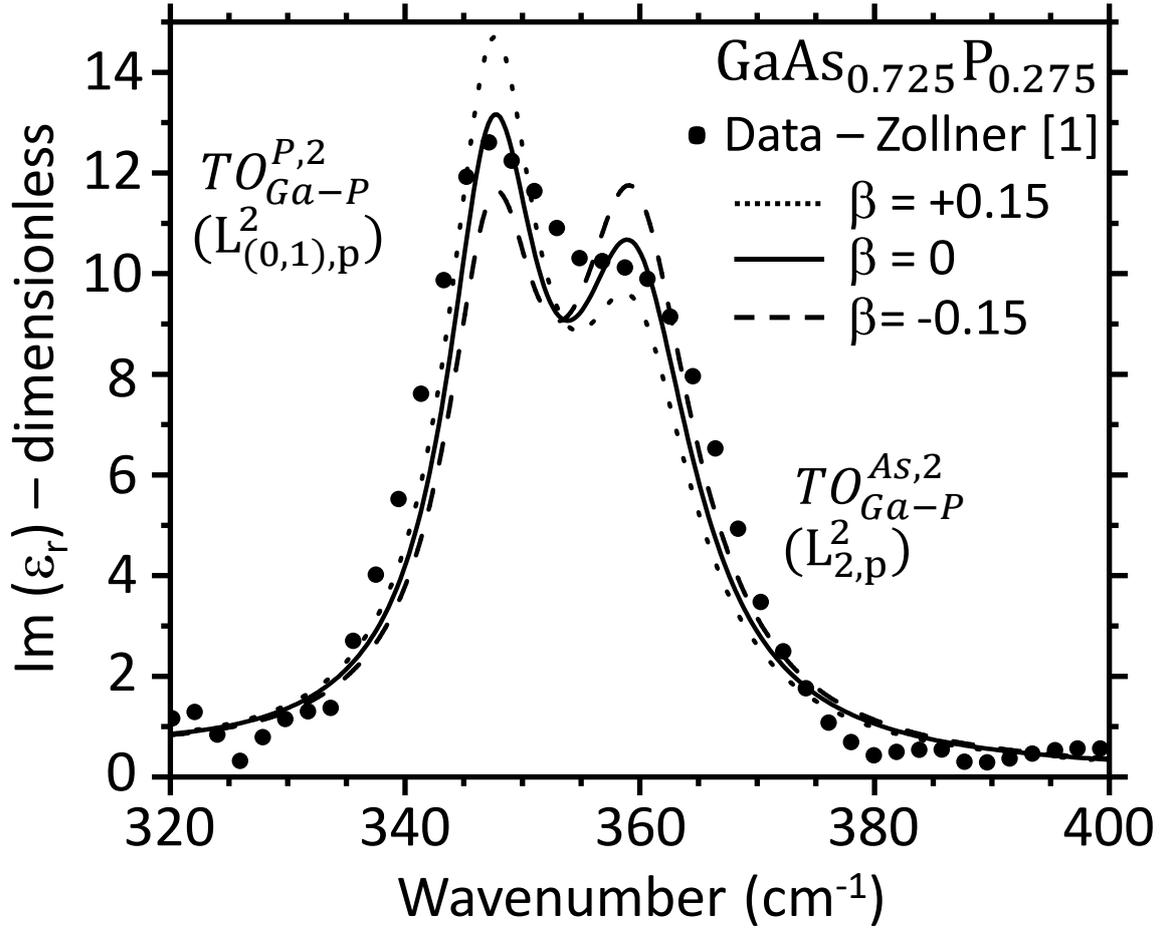

**Fig. S6: $\beta$-dependent GaP-like GaAs$_{1-x}$P$_x$ $Im\{\varepsilon_r(\omega,x)\}$ spectra ($x=0.275$).** Theoretical GaP-like $Im\{\varepsilon_r(\omega, x = 0.275)\}$ GaAs$_{1-x}$P$_x$ spectra obtained within the second-neighbor version of the PM in cases of a random As↔P substitution ($\beta$=0, reference plain line, taken from Fig. S1), of a slight clustering ($\beta$=+0.15, dotted line) and of a slight anticlustering ($\beta$=-0.15, dashed line). The used input parameters are otherwise the same for the three reported curves (given in Table S1). The individual TO modes are labeled using the double $TO$-$L$ notation.